\RequirePackage{fix-cm}
\documentclass[smallextended]{svjour3}       
\smartqed  
\usepackage{graphicx}
\usepackage{color, colortbl}
\usepackage{tikz}

\newcommand{\benchmarktool}{\textit{defect4ML}}
\usepackage{multirow}

\usepackage{csquotes}
\usepackage{dingbat}

\usepackage{xcolor} 
\definecolor{LightGray}{gray}{0.9}

\usepackage{tcolorbox}
\usepackage{hyperref}

\begin{document}

\title{Bugs in Machine Learning-based Systems: A Faultload Benchmark}








\author{Mohammad Mehdi Morovati \and
        Amin Nikanjam \and 
        Foutse Khomh \and Zhen Ming (Jack) Jiang
}


\institute{Mohammad Mehdi Morovati
           \and
           Amin Nikanjam
              \and
            Foutse Khomh \at
            SWAT Lab., Polytechnique Montréal, Montréal, Canada\\
            \email{\{mehdi.morovati,amin.nikanjam,foutse.khomh\}@polymtl.ca}\\
           Zhen Ming (Jack) Jiang \at York University, Toronto, Canada \\
           \email{zmjiang@cse.yorku.ca}
}

\date{Received: date / Accepted: date}

\maketitle

\begin{abstract}
The rapid escalation of applying Machine Learning (ML) in various domains has led to paying more attention to the quality of ML components. 
There is then a growth of techniques and tools aiming at improving the quality of ML components and integrating them into the ML-based system safely. 
Although most of these tools use bugs' lifecycle, there is no standard benchmark of bugs to assess their performance, compare them and discuss their advantages and weaknesses. 
In this study, we firstly investigate the reproducibility and verifiability of the bugs in ML-based systems and show the most important factors in each one. 
Then, we explore the challenges of generating a
benchmark of bugs in ML-based software systems and provide a bug benchmark namely \benchmarktool~ that satisfies all criteria of standard benchmark, i.e. relevance, reproducibility, fairness, verifiability, and usability.
This faultload benchmark contains 100 bugs reported by ML developers in GitHub and Stack Overflow, using two of the most popular ML frameworks: \textit{TensorFlow} and \textit{Keras}. 
\benchmarktool~ also addresses important challenges in Software Reliability Engineering of ML-based software systems, like: 1) fast changes in frameworks, by providing various bugs for different versions of frameworks,
2) code portability, by delivering similar bugs in different ML frameworks,
3) bug reproducibility, by providing fully reproducible bugs with complete information about required dependencies and data, and 4) lack of detailed information on bugs, by presenting links to the bugs' origins. \benchmarktool~ can be of interest to ML-based systems practitioners and researchers 
to assess their testing tools and techniques.
\end{abstract}


\keywords{Benchmark, Machine Learning-based system, Software Bug, Software Reliability Engineering, Software Testing}



\section{Introduction}
Recent outstanding successes in applying Machine Learning (ML) and especially Deep Learning (DL) in various domains have encouraged more people to use them in their systems. 
ML-based systems refer to software systems that contain at least one ML component (software component whose functionality relies on ML).
Given the increasing deployment of ML-based systems in safety-critical 
areas such as autonomous vehicles~\cite{pei2017deepxplore,ma2018deepgauge} and healthcare systems~\cite{esteva2019guide}, we need to provide an acceptable level of reliability in such systems. 

Software reliability is broadly considered to be the most important software quality factor among all software quality attributes~\cite{lyu2007software}, where such attributes measure the conformance level of the system, component, or process to the identified functional and non-functional requirements~\cite{pressman2005software}. Software Reliability Engineering (SRE) is the methodology to ensure failure-free operations of the software in a specified period of time. Substantial portion of SRE techniques has been developed based on studying the lifecycle of bugs~\cite{radjenovic2013software,lyu2007software}.
 
It is generally accepted that standardized benchmarks are the most efficient tools for evaluating and comparing products and methodologies~\cite{v2015build}.
A benchmark must satisfy some quality criteria to be considered as standard, including relevance, reproducibility, fairness, verifiability, and usability~\cite{v2015build,vieira2012resilience}. It is also worth mentioning that benchmark construction is a long-term and iterative process that requires the cooperation of the community~\cite{lu2005bugbench}. 
Accordingly, a standard benchmark of bugs is an essential requirement 
to evaluate, compare, and improve such research on the SRE approaches focusing on the bug's lifecycle. 

Benchmark of software bugs that contains a set of real bugs is known as faultload 
benchmark~\cite{vieira2012resilience}. 
Several studies on the faultload benchmark for traditional software systems have been done, e.g., \textit{Defects4J}~\cite{just2014defects4j} (a benchmark of bugs in Java open source projects hosted on GitHub), \textit{BugBench}~\cite{lu2005bugbench} (a benchmark of C/C++ programs' bugs), \textit{ManyBugs}~\cite{le2015manybugs} (a benchmark of defects in C programming language), 
and \textit{Bears}~\cite{madeiral2019bears} (a Java bug benchmark for automatic program repair). 
On the other hand, some faultload benchmarks such as JaConTeBe ~\cite{lin2015jacontebe} (a benchmark of java concurrency bugs) are designed for specific types of bugs. Accordingly, \benchmarktool also ignores general bugs and considers bugs that are related to the ML components.

Similar to other faultload benchmarks, benchmark of ML-based systems’ bugs is the basic necessity for comparing, tuning, and improving testing techniques/tools of ML-based systems. 
Extracting, reproducing, and isolating real bugs in traditional software
still need considerable time and effort. 
Concerning the higher complexity level of the ML-based systems in comparison with traditional ones~\cite{amershi2019software} and challenges in the engineering of ML-based systems~\cite{galin2004software},
providing reproducible bugs in these systems might require more effort.
Although preceding studies have developed some benchmarks of bugs in ML-based systems~\cite{kim2021denchmark,wardat2021deeplocalize}, they totally disregarded the standard benchmark criteria. As an example, \textit{Denchmark}~\cite{kim2021denchmark} does not provide enough information to reproduce and trigger bugs. 
Meanwhile, 
several studies on the testing of ML-based systems have used synthetic bugs for assessment~\cite{nikanjam2021faults,nikanjam2021automatic} which may bias their evaluation by hiding potential weaknesses.
Some others have also used a limited number of real bugs~\cite{wardat2021deeplocalize,schoop2021umlaut} that may not be representative of a thorough evaluation, implying an incorrect measure of the proposed approach's reliability. 
So, in this research we aim to answer the following research questions:\\
\textbf{\textit{RQ1.}} \textit{What are the key factors in reproducibility of reported bugs in ML-based systems?}
\\
\textbf{\textit{RQ2.}} \textit{What are the important factors in verifiability of ML-based systems' bug-fixes?}
\\
\textbf{\textit{RQ3.}} \textit{What are the challenges of generating standard faultload benchmark in ML-based systems?}

To answer these research questions, 
firstly, we investigate 5 public datasets of ML-based systems' bugs and manually check 513 and 498 bugs that they provided from GitHub and SO, respectively. Then, we checked $1264$ additional bug-fix commits extracted from ML-based systems repositories.
Furthermore, we review $798$ Stack Overflow (SO) posts related to \textit{TensorFlow}~\cite{abadi2016tensorflow} and \textit{Keras}~\cite{chollet2018keras} frameworks. 
We examine the reproducibility and verifiability of the bugs in ML-based systems, as two of the most demanding criteria of standard benchmark.

\begin{figure}
    \centering
    \includegraphics[scale=0.35]{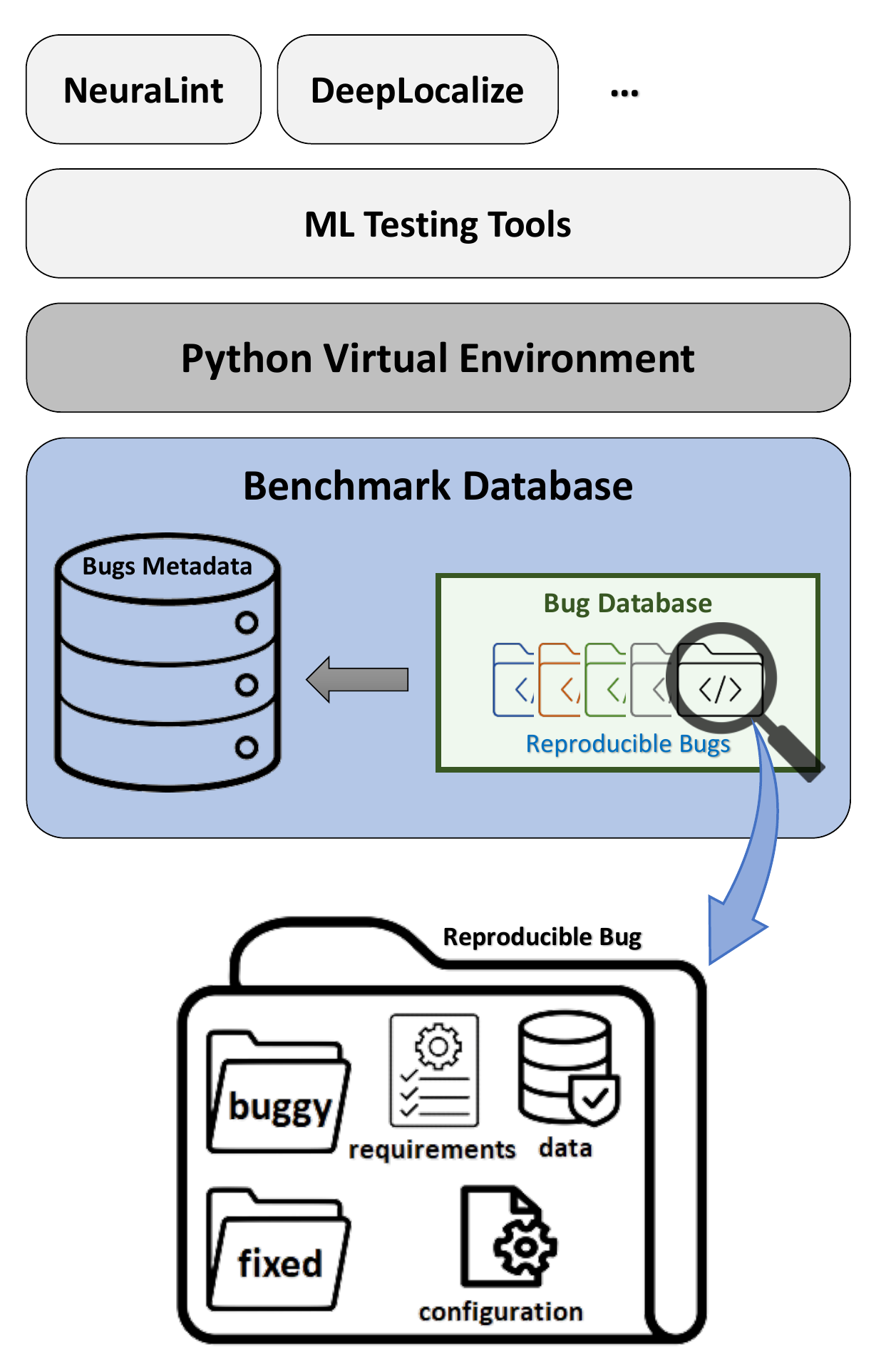}
    \caption{High-level view of the benchmark.}
    \label{fig:benchmark}
    \vspace{-2em}
\end{figure}

We also provide a faultload benchmark of ML-based systems namely \benchmarktool.  
Figure~\ref{fig:benchmark} illustrates a high-level view of the proposed benchmark.
The base layer is the benchmark containing the database of bugs. The next layer, \textit{Python} virtual environment, refers to the environment that should be configured to run applications and trigger the bugs. Because each buggy application has different dependencies and requires different sorts of libraries to be triggered, a \textit{Python} virtual environment would be the best solution for running buggy applications in isolation. The top layer represents the potential usage of bugs which is mostly ML testing tools such as \textit{NeuraLint} and \textit{DeepLocalize}.
Given the high cost of providing bugs from ML-based systems that satisfy standard benchmark criteria, we set 100 bugs (62 from GitHub and 38 from SO) as our goal for the first release of \benchmarktool.
The included bugs are classified into different categories, based on various criteria including Python version, ML framework, violated testing property, and bug type. Different users such as developers, distributors, and researchers of ML-based systems testing tools/techniques can benefit from our proposed benchmark. They can use \benchmarktool~to evaluate their proposed approaches for bug detection, or localization and compare them with previous works. Besides, \benchmarktool{} has potential to be used for automatic bug repairing tools. To this end, users should remove modifications from bug-fix which are not related to the reported bug (similar to methodology used in \cite{jiang2021extracting}).The data generated/analysed during the current study are available in the benchmark repository\footnote{\url{http://defect4aitesting.soccerlab.polymtl.ca/}}.
 
The contributions of this study can be summarized as follows: 
\begin{itemize}
    \item \textbf{First standard 
    benchmark of ML-based systems' bugs}: 
    Proposed benchmark satisfies all of the standard benchmark criteria (e.g. relevance, fairness, etc)
    \item \textbf{Large-scale and accurate
    bug benchmark}: 
    We collected the bugs from GitHub commits and SO posts to provide a large-scale benchmark. Besides, we applied several steps (including manual checking) to filter the bugs and extract the ones satisfying our defined criteria. 
    We have also provided fine-grained classifications based on different criteria (such as ML framework, bug type, etc.) to enable users to filter the bugs and collect a desired subset. Users can also add new bugs to the benchmark and raise a request for removing an existing bug to keep the benchmark up-to-date.  
    \item \textbf{Bug reproducibility}: 
    Our analysis revealed
    that only about $5.3\%$ of all reviewed GitHub bugs and near to $3.34\%$ of reported bugs in SO posts are reproducible.
    However, all bugs in \benchmarktool{} are completely reproducible. We have also provided contextual information for each bug including the needed version of Python, dependencies (necessary libraries and corresponding version), data, and the process of triggering bugs, to allow for reproducibility.  
    
    \item \textbf{Bug-fix verifiability}:
    Our analysis revealed that only a small portion of the studied ML-based systems' bugs (i.e., $13.3\%$ of collected bugs from GitHub in \benchmarktool{}) can be verified by the provided test cases in their applications. 
    Moreover, none of the reviewed bugs reported in SO posts has test cases.
    
    \item \textbf{Detailed information of the bugs}: 
    We provide 
    the URL to the origin of gathered bugs in our proposed benchmark that includes textual information about bugs including: buggy entities such as file name and line of code, bug's root cause, and the fixed version (how the bug got fixed).
    
    \item \textbf{Diversity}: We have covered $30$ different types of bugs based on the taxonomy proposed by Humbatova et al.~\cite{humbatova2020taxonomy} (including 95 types of ML-related bugs), to promote diversity of \benchmarktool. Besides, we used GitHub and SO as the primary sources of collecting bugs. To gather bugs from GitHub, we have also explored repositories developed by users with different levels of expertise. In addition, we have presented bugs based on the two most popular ML frameworks, \textit{TensorFlow} and \textit{Keras}~\cite{top_ml_frameworks,humbatova2020taxonomy}. 
    
\end{itemize}

\textbf{The rest of the paper is organized as follows.} 
We explain the background of the study in Section~\ref{sec:bckground}. The methodology followed to answer the research questions is explained in Section~\ref{sec:method}.
The results of analyzing collected bugs and the proposed benchmark is described in Section~\ref{sec:results}. 
Section~\ref{sec:discussion} represents the discussion of our study. 
Then, the related works are mentioned in Section~\ref{sec:related-work}. We discuss threats to the validity of this research in Section~\ref{sec:validity}. 
Finally, we conclude the paper in 
Section~\ref{sec:conclusion}.

\section{Background}
\label{sec:bckground}
This section introduces concepts of ML-based systems and SRE in these systems, the general picture of the benchmark, and the criteria it should meet to be considered a standard benchmark. 

\subsection{ML-based System}
\begin{figure}
    \centering
    \includegraphics[scale=0.4]{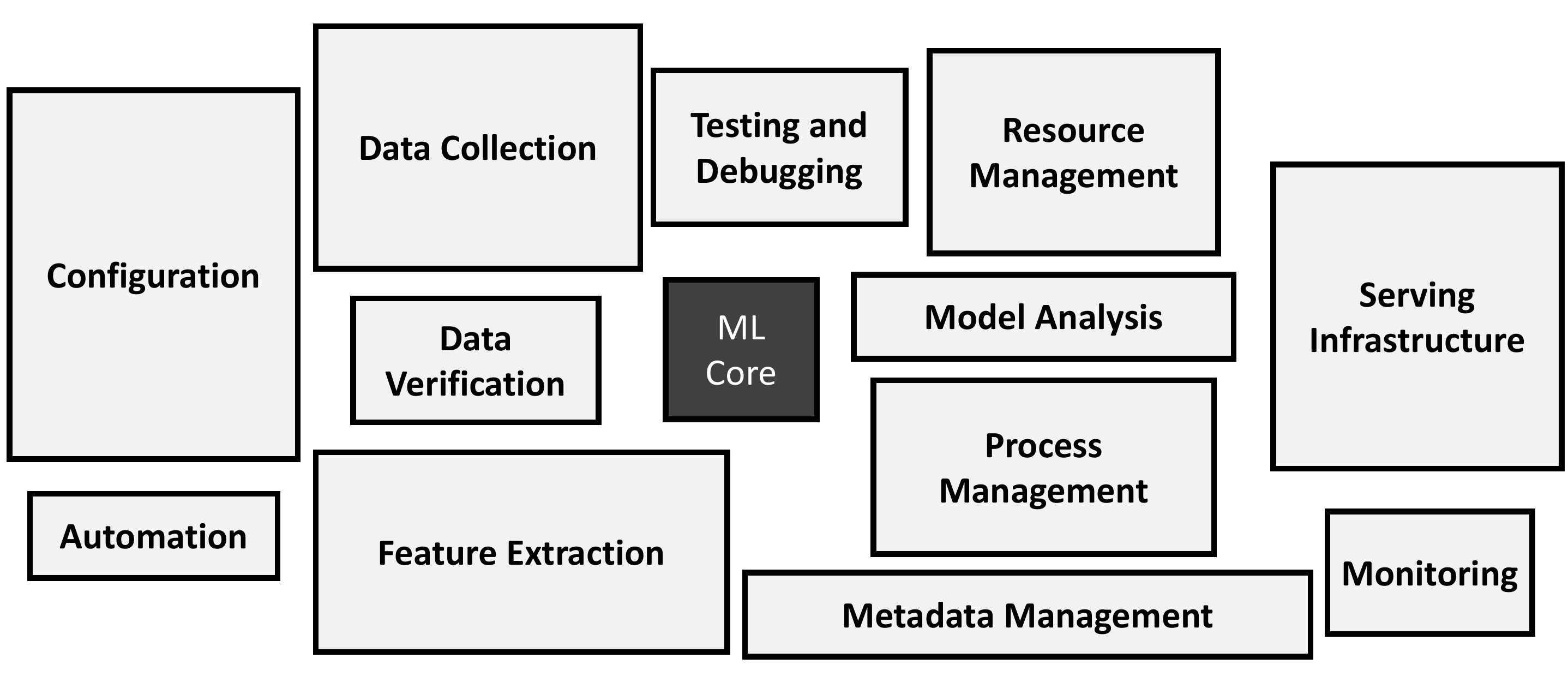}
    \caption{An high-level view of the ML-based systems~\cite{sculley2015hidden}.}
    \label{fig:ml-system}
    \vspace{-2em}
\end{figure}

Software systems including at least one ML component are known as ML-based systems. ML components are defined to be software components working based on ML algorithms with the aim of proving intelligent behavior~\cite{martinez2021software}. 
An ML component may be only a small part of a much larger system. Figure~\ref{fig:ml-system} shows a high-level view of the ML-based systems exposing the role of the ML component in them.

To simplify the design, implementation, and integration of ML components 
in software systems, several
ML frameworks such as \textit{TensorFlow}~\cite{abadi2016tensorflow}, \textit{Keras}~\cite{chollet2018keras}, and \textit{PyTorch}~\cite{paszke2019pytorch} have been developed. 
They help developers to create, train, and then deploy various types of ML models.
Hence, ML frameworks play a vital role in developing ML-based systems~\cite{zhang2020machine}.
Similar to any other software components, ML components are also error-prone. However, current ML frameworks do not provide any capability to validate and verify 
the developed ML components.

\subsection{Bugs in ML-based Systems}
In general, software bug is known as the inconsistency between the existing and expected software functionality, also called deficiency in satisfying software requirements~\cite{zubrow2009ieee}. Accordingly, ML bug refers to the deficiencies in ML components, which may lead to discrepancies between existing and the required behavior of ML component~\cite{zhang2020machine}. 
An ML bug can occur in the ML framework~\cite{jia2021symptoms,rivera2021challenge,tambon2021silent}, program code~\cite{islam2019comprehensive,zhang2018empirical}, or the data. 
So, researchers study the bugs in each area separately~\cite{zhang2020machine}. 
In this study, we just investigate the bugs in program code and do not consider bugs in ML framework and data.

There are three different testing levels for ML-based systems: model testing, integration testing, and system testing~\cite{riccio2020testing}. At the model testing level, we consider the ML component in isolation and ignore other software components. Integration testing level aims to assess the interaction between ML and other components. System testing level studies all software components to evaluate the ML-based system's conformance to the intended requirements. 
Accordingly, we can consider three different categories of bugs in ML-based systems, each one related to one of testing levels. 
\begin{itemize}
    \item Bugs in ML components: At this level, we consider the ML components in isolation and study the bugs inside them ignoring other components of the system. 
    \item Bugs in any component
    that affect functionality of ML components: 
    At this level, all identified bugs in the previous level have been taken into account 
    plus the bugs which are out of the ML components but affect the ML component's functionality. 
    In this paper, we call this category as \textit{\textbf{ML-related}} bugs. 
    \item Bugs in all 
    software components: At this level, we do not make any difference among ML components and others considering all system bugs as the same. 
\end{itemize}

Based on the definition of faults in IEEE standard glossary of software engineering terminology~\cite{ieee5733835:2010}, software fault is manifestation of a bug in software.
In other words, when a software bug causes an incorrect software operation,
it becomes a software fault~\cite{galin2004software}.
Concerning that faults emerge from the discordance between software requirements and the existing behavior, faults can be functional or non-functional. Functional faults refer to the inability of the software to meet the required functionality. Non-functional faults stem from the deficiencies in methodologies to achieve the required functionality, not the functionality.
An ML fault is also considered as an inadequacy in the behavior of the ML component~\cite{humbatova2020taxonomy,ISO21448}. 
A software fault results in a failure only when a user tries to use the faulty software component, 
leading to fault activation~\cite{galin2004software}. 
Generally speaking, failure in software engineering is known as the inability of the system or its components to fulfill required functions~\cite{riccio2020testing}. 
Faults in ML-based systems may also lead to bad performance, crash, data corruption, hang, and memory out of band which are considered as failure in these systems~\cite{islam2019comprehensive}.

\subsection{SRE in ML-based Systems}
\label{subsec:sre_in_ml}

Because of the essential differences between the paradigm of traditional and ML-based software systems, we are facing several new challenges in SRE of ML-based systems. Various studies have acknowledged the significant challenges in SRE of the ML-based systems. 
Fast changes in the new versions of ML frameworks is one of the major challenges that Islam et al.~\cite{islam2020repairing} reported. As an example, they exposed that almost 26\% of operations have been changed from version 1.10 to 2.0 in \textit{TensorFlow}. Code portability is another crucial challenge in the SRE of ML-based systems~\cite{lenarduzzi2021software}. There are multiple ML frameworks (e.g. \textit{TensorFlow}, \textit{Keras}, \textit{PyTorch}, etc.) just for Python programming language. Although they have some similarities, there are major differences among them. Therefore, understanding and porting ML codes from one framework to another can be a nontrivial task. Bug reproducibility is another significant challenge in the SRE of ML-based systems~\cite{zhang2018empirical}. Wardat et al.~\cite{wardat2021deeplocalize} also reported the lack of detailed information regarding the bugs in ML-based systems as a basic challenge in SRE of ML-based systems. We aim to cover these challenges in our proposed benchmark.

SRE techniques mostly use the lifecycle of bugs~\cite{lyu2007software}. 
One of the major SRE approaches using the bug's lifecycle is fault removal that aims to detect the existing faults and remove them. Such techniques use validation and verification approaches to cope with reliability concerns that are known as software testing techniques. Overall, software testing is considered as one of the most complicated tasks of the software development process. It is well-accepted that the complexity of the testing has a direct relation with the complexity level of the system to be tested~\cite{galin2004software}. That means, by increasing the complexity of the system, testing becomes more complicated to be able to deal with the system quality flaws. It has also been proved that the complexity level of the ML-based systems stays at a higher place compared to traditional ones~\cite{amershi2019software}. Consequently, testing of the ML-based systems is considered as more complicated tasks, in comparison with traditional software systems.

ML testing properties refer to the conditions that are needed to be guaranteed for a trained model during testing. In other words, ML testing properties represent the quality attributes that should be tested and satisfied in ML-based systems~\cite{zhang2020machine}. Existence of bugs in ML-based systems may result in violation of various ML testing properties, depending on the impact of bugs on the system. In this study, we used the introduced ML testing properties by Zhang et al.~\cite{zhang2020machine} which are briefly reviewed in this subsection. Correctness represents the probability that the ML system works in the right way, as it is intended. Model relevance checks the complexity of the ML model to make sure it is not more complicated than required. 
In fact, model relevance aims at preventing model overfitting. Overfitting happens when the complexity of the employed ML algorithm is more than required.
Robustness is defined as the extent to which the ML system is able to handle invalid inputs and functions correctly. Efficiency refers to the speed at which ML systems operate and perform the defined tasks (e.g., prediction). Fairness ensures that ML systems make decisions without bias. Interpretability is the degree to which humans can understand the reasons behind the decisions that ML systems make. Although testing properties have been categorized into six different classes, they may overlap with each other~\cite{zhang2020machine}.

Bad performance, crash, data corruption, hang, incorrect functionality, and memory out of bound are symptoms of ML-related bugs which are known as various failure types in ML systems~\cite{islam2019comprehensive,zhang2018empirical}. Bad/poor performance refers to the situation where the accuracy of the ML component is not as good as expected. Crash is the most common symptom of ML-related bugs in which ML-based software stops running with or without showing an error message. Data corruption means data has been corrupted when it passes the network which leads to wrong output. When the ML software stops responding to the input without prompting an error, it is known as hang. Incorrect functionality refers to the situation that ML software behavior differs from the expected, without any error. Memory out of bound occurs due to the unavailability of required memory for training. It should be also taken into consideration that symptoms of bugs belonging to each ML testing property can be different. For instance, if the correctness testing property of an ML component gets violated, its symptoms may be any of known types such as bad performance, crash, hang, etc.

It should be also taken into consideration that symptoms of bugs belonging to one ML testing property can be different. For instance, when the correctness testing property of the ML component has been dissatisfied, its symptoms may be any of known ML-based systems failure types such as bad performance, crash, hang, etc.

It is not surprising that researchers have used testing techniques from traditional software systems to cope with testing challenges of ML-based systems. However, traditional testing methodologies would not be sufficient and efficient testing approaches for ML-based systems~\cite{marijan2019challenges}.
Traditional testing methods require adaptation to the context of ML to be effective for them~\cite{bourque1999guide}. Moreover, the concept of quality is not well-defined in ML-based systems and its terminology is different from the traditional ones~\cite{lenarduzzi2021software,borg2021aiq}.
It is also worth noting that the intrinsic difference between ML-based and traditional software systems generates new types of bugs which do not exist in the traditional software systems~\cite{riccio2020testing}.
For instance, the behavior of the ML-based systems is heavily dependent on factors such as training dataset, hyperparameters, optimizer, etc. Besides, it is 
hardly possible
for humans to debug the learned behavior which is encoded by weights within the ML model.

Several studies have been carried out to provide tools to test ML-based systems. Wardat et al.~\cite{wardat2021deeplocalize} conducted research to localize the bugs in ML-based systems. They explained that because understanding ML models’ behavior is challenging, existing debugging methods for ML-based systems do not support localization of the bugs. They provided a dynamic mechanism to analyze the ML components and implemented an alternative “callback” mechanism in \textit{Keras}
to collect the detailed information of the ML component during the training phase. Then, their proposed tool analyzes the collected data to discover possible bugs and their root causes. 
Islam et al.~\cite{islam2020repairing} carried out an empirical study on the challenges that the automated repairing tools should address. 
They reviewed SO 
posts and GitHub bug fixes using the five most popular ML frameworks (\textit{Caffe}\footnote{https://caffe.berkeleyvision.org/}, \textit{Keras}\footnote{https://keras.io/}, \textit{TensorFlow}\footnote{https://www.TensorFlow.org/}, \textit{Theano}\footnote{https://github.com/Theano/Theano}, and \textit{Torch}\footnote{http://torch.ch/}) to identify fix patterns.
They classified the bug-fix patterns specially used in Deep Neural Networks (DNN) into 15 different categories and provided various solutions to fix bugs
belonging to different classes.
Schoop et al.~\cite{schoop2021umlaut} offered a system
namely UMLAUT to assist non-expert users in identifying, understanding, and fixing bugs in the 
DL programs.
UMLAUT can be attached to the DL program to check the model structure and its behavior.
Then, it suggests the best practices to improve the quality of ML components. 
Nikanjam et al.~\cite{nikanjam2021automatic} provided an automatic fault detection tool for DL programs namely \textit{NeuraLint} that validates the DL programs by detecting faults and design inefficiencies in the implemented models. They identified 23 different rules using graph transformations to detect various types of bugs in DL programs. 

\subsection{Benchmark}
Benchmark is known as a standard tool for the competitive evaluation of systems and for making comparisons amongst systems or components, in terms of specific characteristics like performance, security, and efficiency~\cite{vieira2012resilience}.
It is widely acknowledged that a standardized benchmark is the most significant requirement to evaluate and compare the methodologies~\cite{v2015build}.
To generate a standardized benchmark,
several criteria should be satisfied in the development process of the benchmark, including ~\cite{v2015build,vieira2012resilience}:
\begin{itemize}
    \item \textbf{Relevance:} asserts that the result of benchmark can be used to measure the performance of the operation in the problem domain. That is to say, how the benchmark behavior
    relates to the behavior of interest to its consumers. Relevance is mostly considered as the most important factor of any standard benchmark~\cite{v2015build}. Without providing relevant information to the benchmark users, it is highly possible that the benchmark will not be in the users' interest, even if it gives perfect services for other criteria. As a general rule of thumb, the benchmark that is well-suited for a particular domain has limited applicability, while the benchmark trying to cover a broader range of domains will be less meaningful for any specific domains~\cite{huppler2009art}.
    \item \textbf{Reproducibility:} refers to the benchmark ability to provide the same results while it is run using the same configuration. 
    \item \textbf{Fairness:} explains that all competing systems be able to use the benchmark equally. In other words, the benchmark should be usable for all systems, without generating artificial limitations. 
    \item \textbf{Verifiability:} ensures that the benchmark results are accurate.
    \item \textbf{Usability:} means that the benchmark should be understandable easily to prevent credibility shortage. Credibility shows the level of confidence that users have in the results~\cite{rodriguez2018reproducibility}.
    Ease of use is also another important property belonging to this criterion.
\end{itemize}

Faultload benchmark is one of the main categories of standard benchmark that includes a set of faults and tries to provide experience of the real faults occurring in the system. It is also commonly confirmed that faultload benchmark is the most complex one among all benchmark categories, because of the complicated nature of the faults~\cite{vieira2012resilience}.

Concerning the drastic influence of ML in several safety-critical areas during the last few years, the reliability engineering of ML-based systems has become more crucial. A faultload benchmark of ML-based systems can play a vital role in the assessment of methods working on the reliability engineering of the ML-based systems. 
Although there may exist a great number of benchmarks in each software domain,
a few numbers of them satisfied requirements of the standard benchmark~\cite{vieira2012resilience}. 
Accordingly, there are several public datasets of bugs in ML-based systems provided as either replication package of their study or a faultload benchmark, but they have some considerable problems. 
For example, bug's dataset provided in~\cite{zhang2018empirical}
does not provide any information about the application dependencies and ignores reproducibility of the collected bugs entirely. 
Besides, 
several reported bugs were based on deprecated versions of Python (older than version 3.6) which might be inefficient for assessment of the current ML-based systems testing tools. 

Another public bugs dataset that is provided by Islam et al.~\cite{islam2020repairing} has some similar problems. 
Firstly, they completely disregarded the reproducibility of the bugs. While dealing with the dependency challenge, we came across bugs unrelated to the ML or which occurred in old versions of \textit{Python} (older than 3.6). On the other hand, some mentioned bugs were based on ML frameworks which are discontinued. For instance, this public dataset noted 15 bugs using \textit{Theano} (a \textit{Python} library to define, optimize, and evaluate mathematical expressions)~\cite{al2016theano}. With regards to the fact that \textit{Theano} is a deprecated library and not supported anymore, adding bugs that use \textit{Theano} would not be valuable. 
Also, it reported 17 bugs based on Torch~\cite{collobert2002torch}, a scientific computing framework for ML algorithms, which development has been deactivated since 2018~\cite{torch_readme}.

Similar to the former studies, in the public bugs dataset published by Humbatova et al.~\cite{humbatova2020taxonomy}, many reported bugs depend on \textit{Python} older than 3.6. We also found several commits mentioned as bug-fix, while they are not a real bug~\cite{github_commit_01} or not accessible~\cite{github_commit_02}.
Lack of dependency information is another shortcoming of this public dataset. 

It is worth noting that datasets provided in~\cite{humbatova2020taxonomy,islam2020repairing,zhang2018empirical} are replication packages of those studies and authors do not aim to introduce a faultload benchmark. However, those datasets can be useful for researchers who are studying bugs in ML-based systems.

In the public bugs dataset that Wardat et al.~\cite{wardat2021deeplocalize} have delivered, the reproducibility of the bugs received no attention resulting in many non-reproducible bugs. Besides, the coverage of the provided bugs is also relatively limited. That is to say, they cover limited types of bugs in ML-based systems.

To the best of our knowledge, there is no benchmark for ML-based systems’ bugs that satisfies the mentioned criteria of the standard benchmark. 
\begin{figure*}
    \centering
    \includegraphics[scale=0.41]{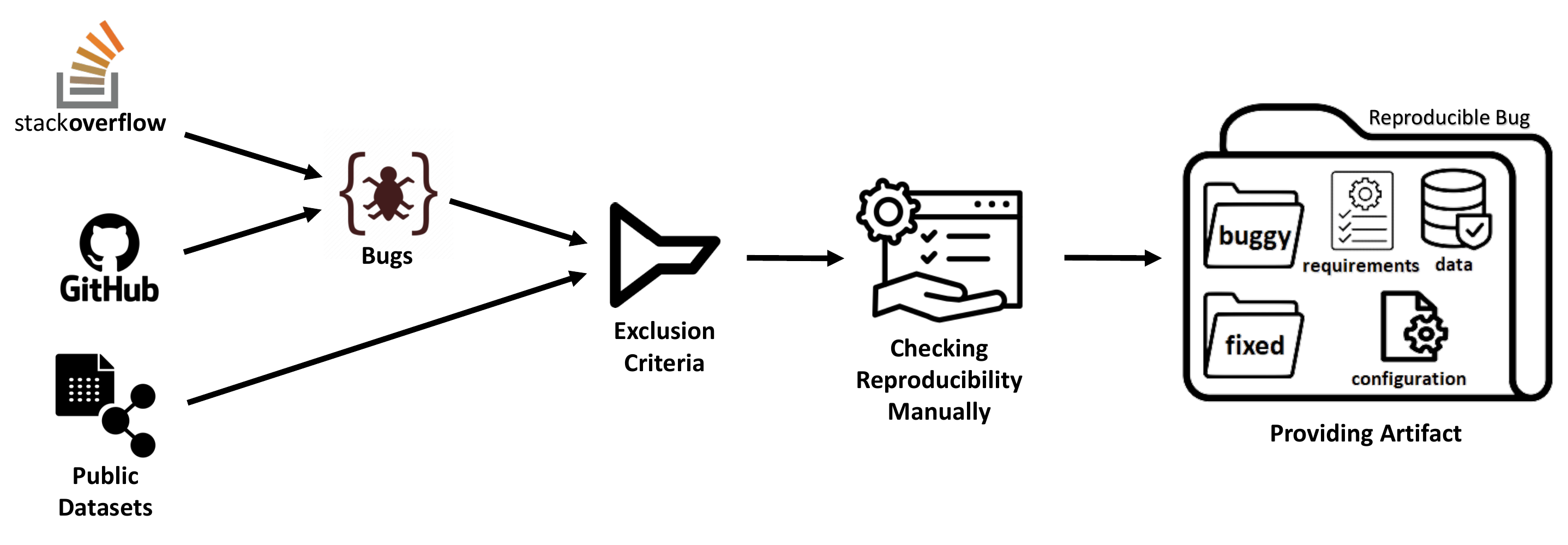}
    \caption{Methodology of collecting bugs.}
    \label{fig:methodology}
    \vspace{-2em}
\end{figure*}

\section{Methodology}
\label{sec:method}
In this section, we describe the methodology that we followed to answer our RQs. 
To this end, we need to collect and investigate the ML-related bugs. 
Figure~\ref{fig:methodology} represents the methodology that we used to collect the bugs for answering our RQs.

We used two main sources to gather bugs: 
(1) public datasets of previous studies on the bugs in ML-based systems, and 
(2) ML-related bugs reported in GitHub or SO. 
To gather the bugs from the prior
research, we reviewed several articles that were about
bug's lifecycle in ML-based systems and provided public bugs datasets~\cite{zhang2018empirical,islam2020repairing,wardat2021deeplocalize,humbatova2020taxonomy}.
As the second source, we extracted the bugs reported in (a) bug-fix commits of GitHub repositories or (b) SO posts. We focused on two of the most popular ML frameworks, \textit{TensorFlow} and \textit{Keras}~\cite{top_ml_frameworks,humbatova2020taxonomy}, respecting the well-known popularity metrics~\cite{zerouali2019diversity} (e.g., number of stars and number of forks) of their GitHub repositories. Table~\ref{tab:ml_frameworks} represents the detailed information regarding the popularity metrics of the selected ML frameworks (on the date we checked them). 

\begin{table}[]
\caption{Detailed information about the selected ML frameworks}
    \centering
    \begin{tabular}{||p{3.5cm}|c|c|c||}
        \hline
            \textbf{ML Framework} & \textbf{\#stars} & \textbf{\#forks} & \textbf{\#subscribers} \\
            \hline\hline
            TensorFlow & $160 K$ & $85.7 K$ & $8 K$\\
            \hline
            Keras & $52.2 K$ & $18.8 K$ & $2 K$\\
        \hline
    \end{tabular}
        \label{tab:ml_frameworks}
\end{table}

We narrowed down our study to the bugs in DL systems, based on the main categories of DL systems faults' taxonomy provided by~\cite{humbatova2020taxonomy}: model, tensors and input, training, GPU usage, and API categories. \enquote{Model} refers to the bugs related to the structure of the model (such as~\cite{github_commit_correctness}). \enquote{Tensor and input} category covers the bugs regarding the problems in the shape and format of data (such as~\cite{so_post_04}). \enquote{Training} includes the bugs in the model training process (such as~\cite{github_commit_efficiencys}). \enquote{GPU usage} category deals with bugs occurred when using GPU devices in DL systems (such as \cite{github_commit_GPU}). \enquote{API} refers to the problems related to the API usage in ML frameworks (such as \cite{github_commit_API}). 

To examine the collected bugs 
and remove those which could not satisfy the standard benchmark criteria in manual checking step, we used some exclusion criteria:
\begin{itemize}
    \item \textbf{bugs using Python version older than $3.6$.} The reason behind this criterion is removing applications using deprecated version of Python or ML frameworks. 
    \item \textbf{bugs irrelevant to the identified ML frameworks (\textit{TensorFlow} and \textit{Keras}).} Some of the collected buggy applications use ML frameworks other than our favorable (e.g., \textit{Caffe}). They are collected because their repositories  have had ``tensorflow'' or ``keras'' keywords in their description. So, we remove them using this criterion. 
    \item \textbf{bugs without dependency information.} One of the most significant needed information to reproduce bugs is their dependencies. We applied this criterion to filter out bugs that are not reproducible. 
    \item \textbf{bugs without required data or description for achieving it.} Used data while facing the bug is another key requirement for reproducing ML-based systems' bugs. We establish this criterion to delete irreproducible bugs.
    \item \textbf{bugs which are not related to the ML.} With respect to our primary aim to investigate characteristics of ML-related bugs and provide faultload benchmark of ML-based systems, we use this criterion to exclude bugs that their root cause is not ML. 
    \item \textbf{bugs with description in any language other than English.} Because we use commit messages as material to inform the users about the detailed information of bug and the bug-fix solution, we deleted the commits that use other languages than English for their commit messages. 
\end{itemize}

\subsection{Collecting Bugs from Public Datasets}
\label{sec:replication_package}
Zhang et al.~\cite{zhang2018empirical} carried out a study on the bugs in \textit{TensorFlow} programs to identify the root causes and symptoms of various types of bugs. They provided a public dataset of 88 and 87 bugs studied in their paper, extracted from GitHub and SO, respectively. After applying exclusion criteria, we obtained 9 GitHub related bugs to be added to the benchmark. But none of their reported bugs from SO remains after applying exclusion criteria.
Islam et al.~\cite{islam2020repairing} provided a public dataset along with their research on the bug fix patterns in DNN programs. DNN program refers to the DL model and training algorithm, where they investigated for bug fix patterns. 
Their public dataset contains 347 and 320 bugs from GitHub and SO, respectively. By filtering bugs using the exclusion criteria, we obtained 8 reproducible \textit{GitHub} related bugs and 3 SO related. 
Wardat et al.~\cite{wardat2021deeplocalize} provided a benchmark of bugs in DNN programs using \textit{Keras}. They reported 11 and 29 bugs from GitHub and SO, respectively. After applying our refinement process, we obtained 4 \textit{GitHub} related and 14 SO related bugs to add to the benchmark. Humbatova et al.~\cite{humbatova2020taxonomy} studied different types of bugs in DL programs and proposed a taxonomy on the identified bugs. They also published the dataset of their studied bugs containing 60 bugs from GitHub and 109 from SO. Our filters eliminated all their mentioned bugs and we could not achieve any bug from their dataset. 
Nikanjam et al.~\cite{nikanjam2021automatic} delivered a public dataset with their automatic bug detection tool including 34 real bugs in DL programs, 26 from SO and 8 from GitHub. After checking their provided bugs using mentioned exclusion criteria, we added 8 of \textit{GitHub} bugs and 10 SO ones to the benchmark. 

Table~\ref{tab:bug_sources} represents the number of bugs that we added to \benchmarktool{} from public datasets of bugs. It is also worth noting to mention that some SO bugs are repetead in different public datasets. As an example, post~\cite{so_post_01} exists in three public datasets~\cite{nikanjam2021automatic,wardat2021deeplocalize,islam2020repairing}.

\begin{table}[]
\caption{Detailed information regarding the collected bugs from public datasets}
    \centering
    \begin{tabular}{l r r}
        \hline
            \multirow{2}{15em}{\textbf{Dataset}} &  \multicolumn{2}{ c }{Num of Collected bugs} \\
             \cline{2-3}
            & \textbf{GitHub} & \textbf{SO}\\
            \hline
            \multicolumn{1}{l}{Zhang et al.~\cite{zhang2018empirical}} & 9 & 0\\
            \multicolumn{1}{l}{Islam et al.~\cite{islam2020repairing} }& 8 & 3 \\
            \multicolumn{1}{l}{Wardat et al.~\cite{wardat2021deeplocalize}} & 4 & 14\\
            \multicolumn{1}{l}{Humbatova et al.~\cite{humbatova2020taxonomy}} & 0 & 0\\
            \multicolumn{1}{l}{Nikanjam et al.~\cite{nikanjam2021automatic}} & 8 & 10\\
            \hline
            \multicolumn{1}{l}{
            \textbf{\textit{Total}}} &\textbf{\textit{29}}& \textbf{\textit{27}} \\
        \hline
    \end{tabular}
        \label{tab:bug_sources}
\end{table}

\subsection{Collecting Bugs from GitHub}
GitHub\footnote{https://github.com/} is considered the most significant resource of open source software repositories in the computer programming community.
As of September 2020, GitHub hosts more than 56 million users and about 190 million software repositories~\cite{github} including more than 28 million public repositories. 
GitHub provides API to simplify the data extraction process that allows developers to create their own requests and extract preferred data. We also used GitHub rest API v3~\cite{github_api_v3} to gather repositories.

\subsubsection{Selection of ML-based Systems Repositories}
To collect the ML-based systems' repositories, we used GitHub search API.
Firstly, we limited the results to the repositories that use Python programming language which is defined as \textit{``Python''} and \textit{``Jupyter Notebook''} programming languages in GitHub. 
\textit{Python} is the most popular programming language for ML~\cite{voskoglou2017best,Gupta:MLLangugae}.
On the other hand, \textit{Keras} and \textit{TensorFlow} also provide \textit{Python} APIs.
As it is mentioned, we analyzed ML-based systems that use \textit{TensorFlow} and/or \textit{Keras}. We extracted the repositories using each of \textit{TensorFlow} and \textit{Keras} separately.
To this end, 
we used ``tensorflow'' and ``keras'' keywords to extract the repositories using these ML frameworks.
In the next step, we limited the repositories to the ones with at least one push after 2019. 
This criterion is to decrease the possibility of reviewing repositories using the old versions of ML frameworks or \textit{Python}, which may not be beneficial to add to the benchmark. At the same time, since it does not prevent the inclusion of repositories using old versions of \textit{Python} or ML framework, we also filter those repositories during our manual inspection. 
Furthermore, repositories that are forked or defined as \enquote{disabled} are excluded in the search query. 

GitHub search API limits the users to access just the first 1000 results. 
So, we divided the whole duration of search for push command from Jan 1, 2019 to
Aug 30, 2021 (the time we run the queries) into snapshots of 5 days to restrict the number of results to less than 1000. That is to say, we raised 192 search requests for each ML framework to extract repositories. We collected $30,387$ and $51,151$ repositories that use \textit{Keras} and \textit{TensorFlow}, respectively. 
In the filtering process, 
we did not filter the repositories based on the popularity criteria (e.g., number of stars, number of commits, etc.) to keep more diversity in the collected bugs. In other words, we aimed to collect
as much as bugs from developers with various expertise levels to avoid generating the benchmark using a biased set of bugs. 

\begin{table}
\caption{
Detailed information about the number of remaining bug-fix commits after each filtering step}
    \centering
   
    \begin{tabular}{p{4.5cm} r r}
        \hline
            \multirow{2}{6em}{} & \multicolumn{2}{ c }{ML frameworks} \\
             \cline{2-3}
            & \multicolumn{1}{c}{\textbf{\textit{TensorFlow}}} & \multicolumn{1}{c}{\textbf{\textit{Keras}}} \\ 
            \hline
            All extracted ML-based repos &  51151 & 30387  \\
            Bug-fix commits & 157190  &  98562 \\
            ML-related bug-fix commits & 38463 & 26326 \\
            Sampled bug-fix commits & 380 & 379 \\
            \hline
            \hline
            \textbf{Added bugs to the benchmark} & \textbf{17}  & \textbf{29}  \\
            \hline
    \end{tabular}
        \label{tbl:GH_bugs}
        \vspace{-1em}
\end{table}

\subsubsection{Selection of Bug-fix Commits}
To extract bug-fix commits from the collected repositories, we searched commits' messages for a list of bug-related keywords (\textit{bug, fail, crash, fix, resolve, failure, broken, break, error, hang, problem, overflow, issue, stop, etc.}), which are used successfully in the literature~\cite{abidi2019anti,abidi2019code,abidi2021multi}. We also used PyDriller~\cite{Spadini2018}, a python library to mine the GitHub repositories, to collect bug-fix commits. In this step, we collected $157,190$ bug-fix commits from repositories using \textit{TensorFlow} and $98,562$ from the ones using \textit{Keras}. 
To exclude bug-fix commits which are irrelevant to ML, we performed another filtering step based on the approach used successfully in~\cite{humbatova2020taxonomy}. 
We searched a list of keywords that are related to the various bug types in ML components (e.g., \textit{optimize, loss, layer}, etc.) in the commits' messages and exclude ones with none of those keywords. Thus, we reached $38,463$ and $26,326$ bug-fix commits for \textit{TensorFlow} and \textit{Keras}, respectively. 

Afterwards, we used sampling with 95\% for confidence level and 5\% for confidence interval that gives us $380$ bug-fix commit for repositories using \textit{TensorFlow} and $379$ for \textit{Keras}.  
In the next step, we manually checked all of the gathered bug-fix commits, applied the exclusion rules, and
removed the inappropriate ones. 
To achieve exclusion criteria, the first author reviewed 200 bug-fix commits and shared the results with the second and third authors (all with more than 2 years of experience in engineering ML-based systems). After three meetings, we achieved an agreement on the following exclusion criteria:
\begin{itemize}
    \item Bug-fix commits with messages written in languages other than English.
    \item Bug-fix commits that do not demonstrate the problem clearly.
    \item Bug-fix commits which used ML frameworks other than \textit{TensorFlow} or \textit{Keras}.
\end{itemize}
Besides, based on the changed LOC and manipulated APIs by the commit, we consider the bug-fix commits that can be categorized as one of the most recent taxonomy of DL bugs~\cite{humbatova2020taxonomy}. To identify ML-related bug-fix commits, the first two authors separately checked the 100 randomly selected bug-fix commits, 50 from repositories developed using TensorFlow and 50 by Keras, attaining 54.7\% agreement using Cohen’s Kappa~\cite{mcdonald2019reliability}. So, we had two meetings to recognize the main reasons for disagreements and resolve them. Next, we again checked the 100 reviewed bug-fix commits achieving 89.6\% agreement based on Cohen’s kappa which is considered as almost perfect agreement~\cite{mchugh2012interrater}. Afterward, the first author reviewed the rest of randomly selected bug-fix commits. Then, the second author checked those bug-fix commits that gained a level of agreement of 86.4\%. 
Concerning most of the reviewed repositories do not provide complete information of their dependencies, we tried to find the match version of the used ML framework with the date of bug-fix commit. For instance, it is obvious that commits which are done before March 5, 2018 
could not use \textit{Keras} higher than version $2.1.4$, because the release date of \textit{Keras} 2.1.5 is March 6, 2018~\cite{keras_repo__2_1_5}. 
Afterward, we attempted to find out the Python version and complete list of required libraries and their version, matching the version of used ML framework.
Finally, we achieved 38 bug-fix commits which meet benchmark requirements. 
Then,
we continued checked manually $505$ other bug-fix commits from repositories using \textit{TensorFlow} and/or \textit{Keras} randomly to attain our goal. Table~\ref{tbl:GH_bugs} represents the number of bugs that we added to the \benchmarktool{} from GitHub.

With respect to the need for both buggy and fixed versions of the application for each identified bug in our proposed benchmark, we have provided a snapshot of the application's repository exactly before fixing the bug. To this end, we use 
\indent\texttt{git log -p <fileName>}\indent
command and extract the commit prior to the bug-fix (
\indent\texttt{fileName}\indent
refers to the buggy file that bug-fix commit will change). Using that commit, we can gather the version of repository exactly before fixing reported bug.  

\subsection{Extracting Bugs from Stack Overflow}
Stack Overflow\footnote{https://stackoverflow.com/} (SO) is taken into account as the largest Q\&A platform for software developers, with over 21 million questions and 14 million registered users on March 2021~\cite{so_insight}. 
To collect intended posts, we used Stack Exchange Data Explore\footnote{https://data.stackexchange.com/stackoverflow/query/new} platform, 
where one can gather information regarding SO posts using SQL queries. 
Like for data extraction from GitHub, we used some criteria for filtering SO posts and collecting relevant ones. In the first step, we collected posts that have ``tensorflow'' or ``keras''
as post tags.
So, we gathered $50,001$ and $37,887$ question regarding \textit{TensorFlow} and \textit{Keras}, respectively. Besides, $21,908$ posts have both ``tensorflow'' and ``keras'' tags at the same time. We considered all of them as posts related to \textit{Keras}. 
Then, we filtered out the posts without an accepted answer where one can not make sure of fixing the issue, in the SO posts without accepted answer. So, we reached $18,812$ posts regarding \textit{TensorFlow} and $14,590$ about \textit{Keras}.
Afterward, we used sampling 
with 95\% and 5\% 
for confidence level and confidence interval, respectively.
Thus, we attained $376$ posts related to \textit{TensorFlow} and $374$ for \textit{Keras}.
It is worth noting that instead of selecting the posts randomly, we selected the posts with the highest scores. 
Next, we manually checked all of the collected bugs' root cause to keep ones related to the ML and remove irrelevant ones. 
Similar to the GitHub manual checking step, we used some exclusion criteria to filter out the irrelevant SO posts. To obtain exclusion criteria, The first author analyzed 100 SO posts, 50 related to the \textit{TensorFlow} and 50 regarding \textit{Keras} with the highest score and discussed the results with the second author. After two meetings, we reached the following exclusion criteria:
\begin{itemize}
    \item posts which mentioned conceptual questions about ML/DL components (such as~\cite{so_post_concept_problem}).
    \item posts related to the users questions on developing ML/DL components, not resolving a bug (such as~\cite{so_post_how_to}).
    \item posts that their root causes were not ML and they were just the typical programming mistakes (such as~\cite{so_post_program_mistake}).
    \item posts that do not include the required script to reproduce the bug.
    \item posts without mentioning the employed dataset or a clear description about it(such as~\cite{so_post_no_dataset}). 
\end{itemize}
In the next step, the first author checked the 50 posts with the highest score for each ML framework (a total of 100) to identify relevant ones. In the next step, the second author reviewed them which obtained 89.3\% agreement using Cohen’s kappa. Then, the first author labeled the rest of the SO posts and the second author checked them, gaining 87.4\% agreement.

\begin{table}
\caption{Detailed information about the number of remaining posts after each filtering out step.}
    \centering
   
    \begin{tabular}{p{4.5cm} r r}
        \hline
            \multirow{2}{6em}{} & \multicolumn{2}{ c }{ML frameworks} \\
             \cline{2-3}
            & \multicolumn{1}{c}{\textbf{\textit{TensorFlow}}} & \multicolumn{1}{c}{\textbf{\textit{Keras}}} \\ 
            \hline
            All extracted posts & 50001  & 37887  \\
            Posts with accepted answer & 18821  & 14590  \\
            Sampled posts & 376 & 374 \\
            \hline
            \hline
            \textbf{Added bugs to the benchmark} & 3 & 7 \\
            \hline
    \end{tabular}
        \label{tbl:SO_bugs}
        \vspace{-1em}
\end{table}
From the availability of the required data viewpoint, we can categorize the SO posts into three main categories. First, the posts that mention popular datasets such as MNIST~\cite{lecun1998gradient} or CIFAR-10~\cite{krizhevsky2009learning} (such as~\cite{so_post_02}). We utilized \textit{Keras} datasets \footnote{https://keras.io/api/datasets/} to reproduce bugs belonging to this group. Second, posts that provide the link to achieve the needed data to reproduce the bugs (such as ~\cite{so_post_03}). Third, posts that did not give any description about the required data (such as ~\cite{so_post_04}). To address the required data problem of the posts that belong to this group, we tried to reproduce the bug using popular datasets. In case of inability to reproduce the reported bug with the same root cause or symptom, we excluded the post.

By manual investigating 376 sampled bugs related to \textit{TensorFlow} and 374 regarding \textit{Keras} (750 in total), we achieved 10 reproducible bugs (3 \textit{TensorFlow} and 5 \textit{Keras}) to add to the \benchmarktool{}. Table~\ref{tbl:SO_bugs} depicts the number of bugs that we added to \benchmarktool{} from SO posts. 
We continued manual checking for 48 more randomly selected SO posts regarding \textit{Keras} that concluded 2 bugs to be added to \benchmarktool{}.

\subsection{Labeling Collected Bugs}
To categorize the collected bugs, we used three kinds of labels, firstly based on violated testing property~\cite{zhang2020machine}, secondly bug type, and finally according to symptoms of bugs~\cite{islam2019comprehensive}. Bug type refers to the class of ML bug taxonomy~\cite{humbatova2020taxonomy} to which the reported bug belongs. For the first step, the first two authors held a meeting to discuss ML testing properties and achieve common understanding on each ML testing property. Then, the first two authors labeled the first 25\% bugs reaching 85.7\% agreement based on Cohen's kappa~\cite{mcdonald2019reliability} which is interpreted as almost perfect agreement~\cite{mchugh2012interrater} and allows us to continue labeling the rest of bugs. To assign ML testing property to each bug, we studied commit message or SO post message to understand the property to which bug-fix aims. For instance, ~\cite{github_commit_efficiencys} which is trying to improve the performance of the ML model is categorized as efficiency property. As another example,~\cite{github_commit_correctness} that resolves the problem in the model structure is labeled as correctness. For labeling the rest of bugs, we labeled them in three parts (25\% of bugs in each part) and held a meeting after each part to identify the major reasons of disagreements and resolve them. In the case of disagreements between two raters, they discussed disagreements with the third author which result in labeling all bugs consistently, which is used in previous studies successfully~\cite{shen2021comprehensive}. Finally, we achieved 88.6\% agreement using Cohen's kappa after labeling all bugs.

For the second labeling step, the first two authors labeled the 25\% bugs separately achieving a 58\% agreement based on Cohen's kappa~\cite{mcdonald2019reliability}, which is known as a moderate agreement~\cite{mchugh2012interrater}. Hence, to improve their understanding of bug types, they had a meeting to have a clear knowledge about each label, identify the main reasons of disagreements, and build a common understanding of bug types. Afterward, raters labeled the first 25 bugs again resulting in 87\% agreement, implying an almost perfect agreement between them. The rater continues labeling the rest of bugs with this approach in three parts (every 25\% of the bugs in each part). After labeling bugs in each part, we had a meeting to recognize the main reasons behind the disagreements and resolve them. Finally, we achieved 88.7\% agreement by labeling all bugs, using Cohen’s kappa. For the remaining disagreements, we used methodology mentioned for labeling bugs based on the violated properties.

To label the bugs according to their symptoms, firstly we had a meeting to achieve an obvious understanding about symptom of ML-related bugs. Next, the first two authors labeled 25\% of the bugs gaining 88.5\% agreement that is interpreted as almost perfect agreement~\cite{mchugh2012interrater}. Then, raters labeled the rest of bugs in three parts (similar to two prior steps) reaching 91.4\% agreement using Cohen’s kappa~\cite{mcdonald2019reliability}. To achieve consistent labels for all bugs, raters discussed the disagreements with the third author and resolved them (same as previous steps). 

\section{Results}
\label{sec:results}
Generally, we manually checked 513 and 498 reported ML-related bugs extracted fron GitHub and SO, and provided in previous articles. 
Also, we collected $64,789$ bug fix commits from ML-based systems repositories using \textit{TensorFlow} and/or \textit{Keras}. We selected 1264 out of them randomly and manually checked them as well for satisfaction of standard benchmark criteria. Moreover, we manually inspected 798 SO posts related to \textit{TensorFlow} and/or \textit{Keras}. In the following, we present our answers to each of our formulated research questions.

\subsection{RQ1. Key factors in reproducibility of ML-related bugs}
Bug reproducibility plays a key role in this study, because it is taken into account as a substantial challenge in SRE of ML-based systems on the one hand~\cite{zhang2018empirical} and in the benchmark generating on the other hand~\cite{v2015build}. To make sure of reproducibility of the collected bugs, we did several manual checking steps. Firstly, we checked the buggy application for needed dependencies information to run it without dependency issues.  
In the next step, we looked into needed data for running the buggy application and triggering the reported bug. After addressing dependency and data requirements of each bug, we faced some new challenges while trying to run the buggy applications.  
Regarding the extracted ML-related bugs from GitHub, 
we have found several bugs with compilation errors while they are running (such as~\cite{github_commit_04}). Besides, some of the bugs were not triggered using the mentioned condition in bug-fix commit (such as~\cite{github_commit_05}). 
We coped with the reproducibility problem of 62 out of 1777 (513 from public datasets and 1264 from GitHub) reviewed ML-related bugs extracted from GitHub. 

About the reported bugs in SO, almost all of the scripts mentioned in the posts are just code snippets, not a complete code of the ML component, as it is popular in SO. 
Therefore, we had to complete them by the default configuration and then check to ensure that the considered bug would occur.
In some cases, we failed to use the bug
after spending considerable time. Moreover, most of the SO posts do not include the required dependency and data information to reproduce the reported bugs. We could reproduce 38 out of 1296 (498 from public datasets and 798 from SO) SO posts. 

\begin{tcolorbox}
\textbf{Finding 1: } The most influential factors in reproducing ML-related bugs are required dependencies (including used libraries and their exact version, ML framework version, and Python version) and data to run the buggy application. Besides, only near $3.48\%$ and $2.93\%$ of all manually inspected ML-related bugs extracted from GitHub and SO can be reproduced. 
\end{tcolorbox}

\subsection{RQ2. The important factors in verifiability of fixing ML-related bugs}
Software verification refers to the process of checking software to ensure that it achieves defined goals without any bug~\cite{8055462:IEEE2016}. 
To verify the fixing of reported bugs, we need to ensure that the mentioned problem will be resolved after applying the fix. 
Software systems use software testing methods to verify their goals and objectives~\cite{lyu2007software}. 
We investigated all reproducible ML-related bugs for the test cases provided by buggy applications that check the verification of the software component consisting of reported bugs.
We found 10 out of 75 buggy applications extracted from GitHub that can be verified using provided test cases by the applications. In case the buggy application does not give any test cases, we have to trust the bug-fix commit message to verify the bug fixes, like almost all of the previous articles studying ML-related bugs and their fixes~\cite{humbatova2020taxonomy,zhang2018empirical,islam2020repairing}.

Verifiability of fixing reported bugs in SO posts is a more severe challenge. Most of the scripts mentioned in SO posts are code snippets and do not have additional information like test cases. So, we could not verify any of the reproducible ML-related bugs extracted from SO using provided test cases by developers who ask the question. 
Instead, we used the accepted answer flag to verify bug fixes. 

\begin{tcolorbox}
\textbf{Finding 2:} Lack of test cases and immaturity of providing test cases for ML-related bugs are the most influential factors for verifying fix of ML-related bugs. 
Moreover, just near $13.3\%$ of ML-related bugs gathered from GitHub can be verified using provided test cases by their applications. However, none of the reproducible ML-related bugs from SO could be verified by test cases implemented by the user who created the post. 
\end{tcolorbox}

\subsection{RQ3. Providing a standard faultload benchmark for ML-based systems}
\begin{figure}
    \centering
    \includegraphics[width=0.5\columnwidth]{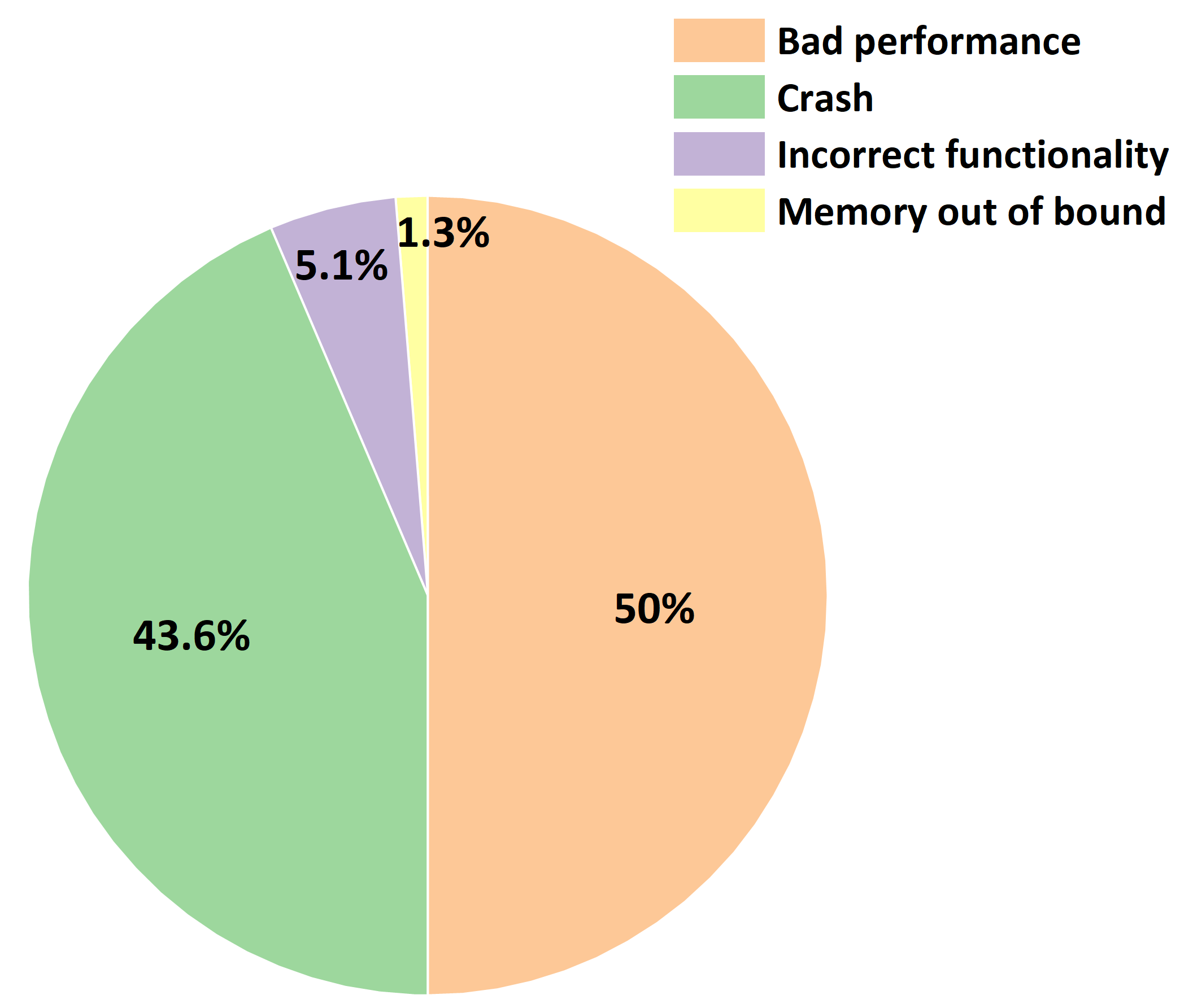}
    \caption{Distribution of bugs’ symptoms in defect4ML.}
    \label{fig:bug_symptoms}
    \vspace{-2em}
\end{figure}

Overall, we collected 100 bugs for \benchmarktool, 62 from GitHub, and 38 from SO. Table~\ref{tab:details_bugs} shows the detailed information of collected bugs. We will explain the challenges we faced while generating the \benchmarktool{} and its merits in comparison with others in the upcoming subsections. Figure~\ref{fig:bug_symptoms} also depicts the the distribution of bugs, based on their symptoms. 
\begin{table}[]
 \caption{Detailed information of bugs.
    }
    \centering
    \scriptsize		
    \setlength\tabcolsep{1.5pt}
\begin{tabular}{|c|c|c|l|c|c|}
        \hline
        \rowcolor{lightgray}
        \textbf{Source} &\textbf{Framework} &\textbf{Bug category}& \multicolumn{1}{|c|}{\textbf{Bug type}}  & \textbf{\#bugs} & \textbf{Total} \\
        \hline
        \multirow{30}{*}{GitHub}&\multirow{13}{*}{TensorFlow}& \multirow{3}{*}{API} & Deprecated API & 2& \multirow{3}{*}{\textbf{6}} \\
        \cline{4-5}
        & & & Missing variable initialization  & 1& \\
        \cline{4-5}
        & & &Wrong API usage & 1& \\
       \cline{3-6}
        & &\multirow{4}{*}{Model} & Missing softmax layer & 1 & \multirow{4}{*}{\textbf{4}}\\
        \cline{4-5}
        & & & Wrong network architecture & 1 & \\
       \cline{4-5}
        & & & Wrong type of activation function & 1& \\
        \cline{4-5}
        & & & Wrong weights initialisation & 1 & \\
       \cline{3-6}
       & & \multirow{2}{*}{ Tensors \& inputs } & Tensor shape mismatch & 1 &\multirow{2}{*}{\textbf{2}} \\
        \cline{4-5}
        & & & Wrong shape of input data & 1 & \\
        \cline{3-6}
        & & \multirow{4}{*}{Training} & Redundant data augmentation & 1 &  \multirow{4}{*}{\textbf{8}}\\
        \cline{4-5}
        & & & Suboptimal learning rate & 2 & \\
        \cline{4-5}
        & & &Wrong loss function calculation  & 4 & \\
        \cline{4-5}
        & & & Wrong selection of loss function & 1 & \\
        \cline{2-6}
         &\multirow{17}{*}{Keras} &\multirow{4}{*}{API} & Deprecated API&  3&\multirow{4}{*}{\textbf{8}} \\
         \cline{4-5}
         & & & Missing API call & 2 & \\
         \cline{4-5}
         & & &Missing argument scoping & 1 & \\
         \cline{4-5}
         & & & Wrong API usage &  2& \\
         \cline{3-6}
         & & \multirow{6}{*}{Model}& Missing dense layer&1  & \multirow{6}{*}{\textbf{14}}\\
         \cline{4-5}
         & & &Suboptimal network structure & 4 & \\
         \cline{4-5}
         & & &Wrong filter size for convolutional layer &  1& \\
         \cline{4-5}
         & & & Wrong layer type & 2 & \\
         \cline{4-5}
         & & & Wrong network architecture& 3 & \\
         \cline{4-5}
         & & &Wrong type of activation function & 3 & \\
         \cline{3-6}
         & & Tensors \& inputs & Wrong tensor shape &  1& \textbf{1}\\
         \cline{3-6}
         & &\multirow{6}{*}{Training} & Missing preprocessing step& 1 & \multirow{6}{*}{\textbf{19}}\\
         \cline{4-5}
         & & &Suboptimal batch size & 4 & \\
         \cline{4-5}
         & & &Suboptimal number of epochs & 4 & \\
         \cline{4-5}
         & & & Wrong loss function calculation& 1 & \\
         \cline{4-5}
         & & &Wrong optimisation function &4  & \\
         \cline{4-5}
         & & &Wrong selection of loss function &  5& \\
         \cline{1-6}
         \multirow{14}{*}{SO} & \multirow{2}{*}{TensorFlow} & API& Deprecated API& 2 & \textbf{2}\\
         \cline{3-6}
         & & Tensors \& inputs& Wrong input format& 1 & \textbf{1}\\
         \cline{2-6}
         &  \multirow{12}{*}{Keras}& \multirow{2}{*}{API}& Wrong API usage&  4& \multirow{2}{*}{\textbf{5}}\\
         \cline{4-5}
         & & & Deprecated API &1 & \\
         \cline{3-6}
         & & \multirow{3}{*}{Model}& Suboptimal network structure& 1 & \multirow{3}{*}{\textbf{11}}\\
          \cline{4-5}
         & & & Missing Flatten layer & 2 & \\
         \cline{4-5}
         & & &Wrong type of activation function &8  & \\
         \cline{3-6}
         & & \multirow{3}{*}{Training}& Suboptimal learning rate&3& \multirow{3}{*}{\textbf{9}} \\
         \cline{4-5}
         & & & Wrong loss function & 2 &\\
         \cline{4-5}
         & & & Missing preprocessing & 4 & \\
         \cline{3-6}
         & & \multirow{4}{*}{Tensors \& inputs}& Tensor shape mismatch& 2& \multirow{4}{*}{\textbf{10}}\\
         \cline{4-5}
         & & &  Wrong type of input data &  1& \\
         \cline{4-5}
         & & & Wrong shape of input data& 6 & \\
         \cline{4-5}
         & & & Wrong tensor shape& 1 & \\
        \hline
        \rowcolor{lightgray}
        \multicolumn{5}{|c|}{\textbf{\textit{Total}}} & \textbf{\textit{100}} \\
        \hline
\end{tabular}
       \label{tab:details_bugs}
\end{table}

\subsubsection{Satisfied Criteria of Standard Benchmark  
}
We consider all criteria of standard faultload benchmark as the challenges of generating standard bug benchmark in ML-based systems. We will describe the methodologies we used to satisfy each criterion in the upcoming subsections. 
\paragraph{\textbf{Relevance.}}
To meet the relevance criterion, we just include ML-related bugs in \benchmarktool. Because there are well-suited benchmarks for traditional programming bugs (non ML-related bugs) such as~\cite{widyasari2020bugsinpy,madeiral2019bears,le2015manybugs,lu2005bugbench,just2014defects4j,lin2015jacontebe}, we 
exclude any bugs that are not related to ML. For example, commit~\cite{github_commit_03} which is related to the fix in \textit{README} file has been excluded from \benchmarktool.

\paragraph{\textbf{Reproducibility.}}

To fulfill reproducibility, we have provided accurate information 
required
for reproducing bugs. The information consists of complete list of dependencies (all needed libraries and their exact version including ML framework), Python version compatible with ML framework and other libraries that application uses, used dataset while facing the reported bug, and the instruction to activate the bug.

\paragraph{\textbf{Fairness.}}
We have applied three main measures to meet the fairness and prevent generating artificial limitations for \benchmarktool.  
Firstly, we have equipped the benchmark with 
two of the most popular ML frameworks (\textit{TensorFlow} and \textit{Keras}~\cite{top_ml_frameworks,humbatova2020taxonomy}). 
Secondly, we have used the taxonomy of bugs in DL programs introduced in~\cite{humbatova2020taxonomy} and tried to label the benchmark bugs based on the leaves of that taxonomy. 
We have covered $30$ types of bugs mentioned in this taxonomy.
So, the benchmark can satisfy various users’ requirements who focus on specific types of bugs. 
We have also provided different types of bugs using various versions of ML frameworks and Python.

\paragraph{\textbf{Verifiability.}}
We have implicitly satisfied this criterion because all collected bugs are real and discussed in GitHub projects or SO posts. 
The benchmark also provides a link to the bug's origin (GitHub bug-fix commit or SO post) for all bugs.
Bug-fix commit messages and SO posts represent detailed information about the occurrence of the bug and the solution to fix it. 
The benchmark delivers two versions of the application where the bug has occurred, i.e., the \textit{buggy} and \textit{fixed} (or clean) versions. The \textit{Buggy} version indicates the version of the application before fixing the bug. The \textit{Fixed} version refers to the application after fixing the identified bug.
Regarding the verifiability of fixing bugs collected for \benchmarktool, 10 out of 62 reproducible ML-related bugs gathered from GitHub are verifiable by the provided test cases and the rest based on the bug-fix commit massage. 
We also used accepted answer flag for SO posts that prove the verifiability of the solution provided to resolve the reported bug. 
\paragraph{\textbf{Usability.}}
We aim to generate an understandable benchmark by delivering information with each bug, such as violated testing property, bug's type, and a link to the bug's origin. 

Users can achieve detailed information regarding the bug's root cause, symptoms, and possible fixing methods using the mentioned link. We have also provided several categorizations (bug's origin, Python version, ML framework, violated testing property, and bug's type)
to prepare a set of bugs that fit users' aims.
Users may use \benchmarktool~ for different goals, such as assessing the ability of testing tools that focus on bug detection, repair, and localization. 

\subsubsection{Addressed SRE Challenges in ML-based Systems}
Several new challenges in the SRE of ML-based systems make it more complicated compared to the traditional software systems~\cite{islam2020repairing,wardat2021deeplocalize}. Neglecting these challenges may deteriorate the satisfaction level of relevance, reproducibility, verifiability, and usability of ML-based systems. 
We rise to these challenges in \benchmarktool~ to ensure the relevance of our proposed benchmark. 
In the following subsections we elaborate on the 
methods used to handle each challenge.

\paragraph{\textbf{Fast Changes in ML Frameworks.}}
To handle the challenge of fast changes in the ML framework, users need to have the exact information regarding the version of the used ML framework in the application containing the bug. We have presented 
this information
for current bugs in \benchmarktool{}\footnote{\label{note1}In \textit{requirements.txt} file, that consists of detailed information of dependencies per bug.}.
Also, \benchmarktool~ has been equipped with bugs that appeared in different versions of each ML framework.

\paragraph{\textbf{Code Portability.}}
To tackle the code portability challenge, defect4ML has provided different bugs that occurred in the applications developed using the two most popular ML frameworks: \textit{TensorFlow} and \textit{Keras}~\cite{top_ml_frameworks,humbatova2020taxonomy}. Therefore, users have access to a list of bugs in their preferred ML framework, without requiring porting bugs from one ML framework to another one. Users can also enhance the benchmark by adding bugs from ML frameworks other than those studied in this paper or recreating bugs in new ML frameworks.

\paragraph{\textbf{Bug Reproducibility.}}
Bug reproducibility is known as one of the most critical challenges in all SRE areas. But it is a more severe challenge in ML-based systems, because of direct effect on the other challenges of SRE in ML-based systems such as fast changes of ML frameworks and code portability.
This difficulty may result from 1) a high amount of operational changes in versions of the ML frameworks~\cite{islam2020repairing}, and 2) different dependencies specified for every single version of ML frameworks~\cite{zhang2018empirical}.
To meet this challenge, we have delivered 
a complete list of
dependencies (and corresponding versions) needed to run the application\textsuperscript{\ref{note1}}.
We have also presented specific configuration of each bug including Python version, and process to trigger the bug\footnote{In \textit{conf.ini} file, that contains the required configuration per bug.}.
Moreover, because of the substantial effect of data in ML components operations~\cite{zhang2020machine,felderer2021quality}, we have delivered the required training/testing datasets 
to reproduce the bugs. 

\begin{figure}
    \centering
    \includegraphics[scale=0.5]{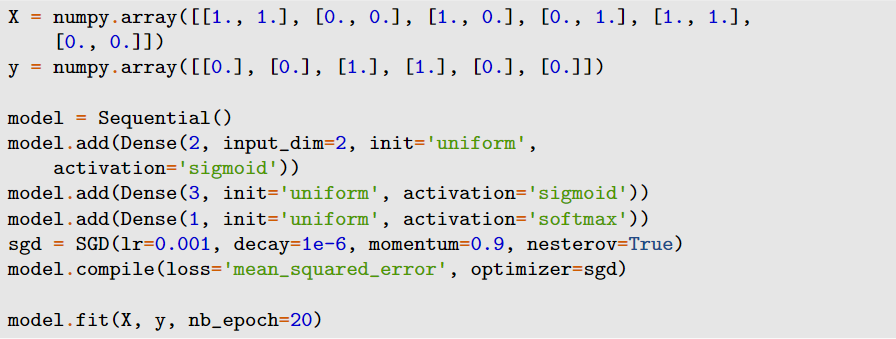}
    \caption{Sample implementation of XOR classification problem using \textit{Keras}~\cite{so_post_01}.}
    \label{fig:code_snippet}
    \vspace{-2em}
\end{figure}

\paragraph{\textbf{Lack of Detailed Information about the Bugs.}}
When an ML component faces a bug, the compiler mostly gives no detailed information regarding the root cause of the bug or the exact bug location.
For example, the compiler may not recognize the ML model structural bugs, or provides just an error message to inform the developer about the existence of a problem in the ML model structure.
But it does not give any clue (such as bug location) to the developer for debugging the ML model.
Figure~\ref{fig:code_snippet} shows a sample script trying to implement a classifier for $XOR$ problem using \textit{Keras}
(bug \href{https://stackoverflow.com/questions/34311586}{\#84} of \benchmarktool{})

~\cite{so_post_01}. Although loss remains constant during training (because of wrong activation function in the last dense layer), compiler does not give any information about the problem to the user.
To cope with this challenge, we use GitHub bug-fix commit messages
and SO posts description that give detailed bugs specifications and the possible solutions to fix them. Bug-fix commit messages may also include the link to the raised issue in the issue tracker, which has in-depth information about the bug. 

\subsubsection{Artifacts}
\label{sec:bug_preparation}
Each bug consists of several components to meet the prerequisites of the benchmark. These components include:
\begin{itemize}
    \item \textit{\textbf{Buggy and fixed versions of the application:}} each bug has two different versions of the application containing that bug.
    \textit{Buggy} version is the application including the bug and is generally used to evaluate ML-based systems testing tools' ability to detect the bugs.     
    \textit{Fixed} version is the same application just after fixing the bug and is mainly used to assess the repairing tools. 
    Repairing tools try to detect the bugs and provide another version of the application where the identified bug has been fixed based on the best practices.
    
    \item \textit{\textbf{Dependencies (Required libraries):}} each bug comes with a complete list of dependencies required to run the {buggy} version (e.g., ML framework and its version, and needed libraries and their versions).
    
    \item \textit{\textbf{Data:}} All bugs are equipped with the needed data to trigger the reported bug. 
    
    \item \textit{\textbf{Configuration:}} 
    that
    is used to produce bug categorization on the benchmark. 
    Moreover, the process of running the application to trigger the bug is mentioned in the configuration. 
    \item \textit{\textbf{Test case:}}
    Each bug has its own test case which can execute the buggy application and trigger the bug, without requiring any user's manipulation. It has been exposed that providing test cases to discover bugs in ML-based systems is more effective than using assertion inside the application~\cite{jia2021unit}. Besides, the test case enables users to observe the effect of the bug and compare the behavior of the system in buggy and fixed versions.
    Because the result of an ML application could be different for each run, even with the same hyperparameters and dataset, providing assertion on exact values can be ineffective for ML application, as shown by Nejadgholi et al. ~\cite{nejadgholi2019study}. Hence, we use a range for assertion of provided test cases in defect4ML. Besides, according to the previous studies showing that accuracy of fixed and buggy version will be close to each other, in case the buggy version does not lead to crash/hang~\cite{jia2022injected}, we use the average result of running buggy and fixed versions 10 times each to address challenges of determining threshold in the provided assertions. 
    \item \textit{\textbf{Detailed description:}} each bug has a detailed description including its root cause, symptom, and an explanation that represents the situation triggering the bug to ease understanding of bugs. 
\end{itemize}


Users who want to use \benchmarktool{} bugs should be informed that we provide test case assertions on that part of results mentioned as symptoms of bug in GitHub issues/SO posts. For example, in bug \href{https://stackoverflow.com/questions/58844149}{\#92}, the user obviously asked for a solution to resolve the problem of low accuracy. Thus, we provide assertions on the accuracy of the buggy and fixed versions. As reported in ~\cite{wardat2022deepdiagnosis}, diagnosing an ML bug may require running the code, collecting information on training and validation phases, and monitoring various values. We are aware that relying on model accuracy to diagnose a bug in ML applications may not always be precise~\cite{10.1145/3324884.3416545}, because of the non-deterministic nature of ML applications leading to different results/accuracy in various executions with the same hyperparameters and dataset. To this end, in cases where a bug is detected based on its impact on the model accuracy, we run each version of an ML application multiple times (mostly 10 times) in the test cases and use the averaged accuracy achieved in multiple executions.

We have classified bugs based on different criteria in ~\benchmarktool. The first criterion is testing properties of ML-based systems~\cite{zhang2020machine} which includes correctness, model relevance, robustness, security, efficiency, fairness, interpretability, and privacy. This criterion refers to the conditions that should be guaranteed for the trained model.

Correctness refers to the ability of a trained model to predict unseen future data correctly~\cite{zhang2020machine}. That is, when an ML model is not designed and/or trained optimally, it can manifest low accuracy during test or deployment. Thus, to meet correctness which requires model accuracy improvement, developers should revise the designed ML model based on recommendations of ML experts. As an example, in the code shown in Figure 5, because of using the wrong activation function (softmax) in the last dense layer, model accuracy stays near 66\%. In fact, the softmax activation function is helpful, where the number of target classes is more than 2. Thus, removing this activation function resolves the problem and increases the model accuracy. All of the currently presented bugs in ~\benchmarktool{} fall into the correctness category.

Model relevance checks for a proper match between the design of the model and training data~\cite{kirk2014thoughtful}. In other words, model relevance asserts that a designed ML model should not be more complicated than what is required. Model overfitting usually leads to low model relevance~\cite{zhang2019perturbed}. Providing a neural network with unnecessary hidden layers may cause model overfitting, and then low model relevance. For example, when a model is more complex and has more learning capacity than needed, it may fit noises of training data resulting in contaminating model generalization~\cite{hawkins2004problem}. Thus, decreasing model learning capacity (adding dropout layer, weight decay, etc ) would improve model generalization and accordingly, model relevance.

Robustness is defined as the degree to which an ML system can handle any perturbation on ML components (e.g. data) ~\cite{ieee:159342,zhang2020machine}. A trained ML model should be able to handle small perturbation on data, like adversarial samples. As an example, studies showed that existence of adversarial examples  in safety-critical systems (such as autonomous vehicles) may lead to significant improvement in the robustness of the system~\cite{tian2018deeptest,zhang2018deeproad}.

Security refers to the resiliency of a ML system against harmful or dangerous actions by illegal access or manipulation of ML components~\cite{xue2020machine}. Systems with low accuracy may deal with data poisoning if perturbed data is employed as training data. A security attack may mislead a trained model or lead a model to be trained badly by manipulating training data~\cite{liu2020privacy}. For example, accessing “stop sign” training data of an autonomous vehicle and manipulating it to decrease detection performance of the model may lead to a catastrophe~\cite{zhu2019transferable}.

Efficiency measures consumed computational resources for training or inferring processes in the ML system~\cite{zhang2020machine}. Overall, an ML system suffering from suboptimal model structure may need more training time compared to the optimal model structure. For example, training an ML model aiming at decreasing loss may be faced with stopping loss decrement after some training loops~\cite{early_stopping_in_dl}. Afterwards, continuing training may be considered a waste of resources. Hence, ML developers use early stopping to monitor evaluation metrics and cease training, whenever training no longer improves evaluation metrics~\cite{rice2020overfitting}.

Fairness aims at preventing ML decisions to suffer from ethical issues (e.g. human rights, discrimination, etc.)~\cite{chouldechova2018frontiers}. In general, human beings have a bias in labeling or collecting data~\cite{10.2307/24758720}. Fairness ensures that ML decisions are in the right way and free of bias. Unfair models may produce discrimination, where they do not work for some subpopulations. An example of an unfair model can be a medical image processing model that works inaccurate, except for white males.

Interpretability refers to the degree that reasons behind decisions made by ML models can be understandable by human beings~\cite{lipton2018mythos}. As an example, an ML model with high interpretability used for medical treatment decisions may be trusted more by medical experts. 
Last but not least, privacy aims at protecting private information that can be used as training data~\cite{dwork2008differential,zhang2020machine}. For example, data used for an ML model that plays a role of assistant in medical treatment decisions should preserve privacy of patients information.

Another filtering criterion is the ML framework used to implement the buggy ML component. To indicate the types of bugs, we used the taxonomy of DL bugs proposed by Humbatava et al.~\cite{humbatova2020taxonomy}.

\begin{figure}
    \centering
    \includegraphics[scale=0.27]{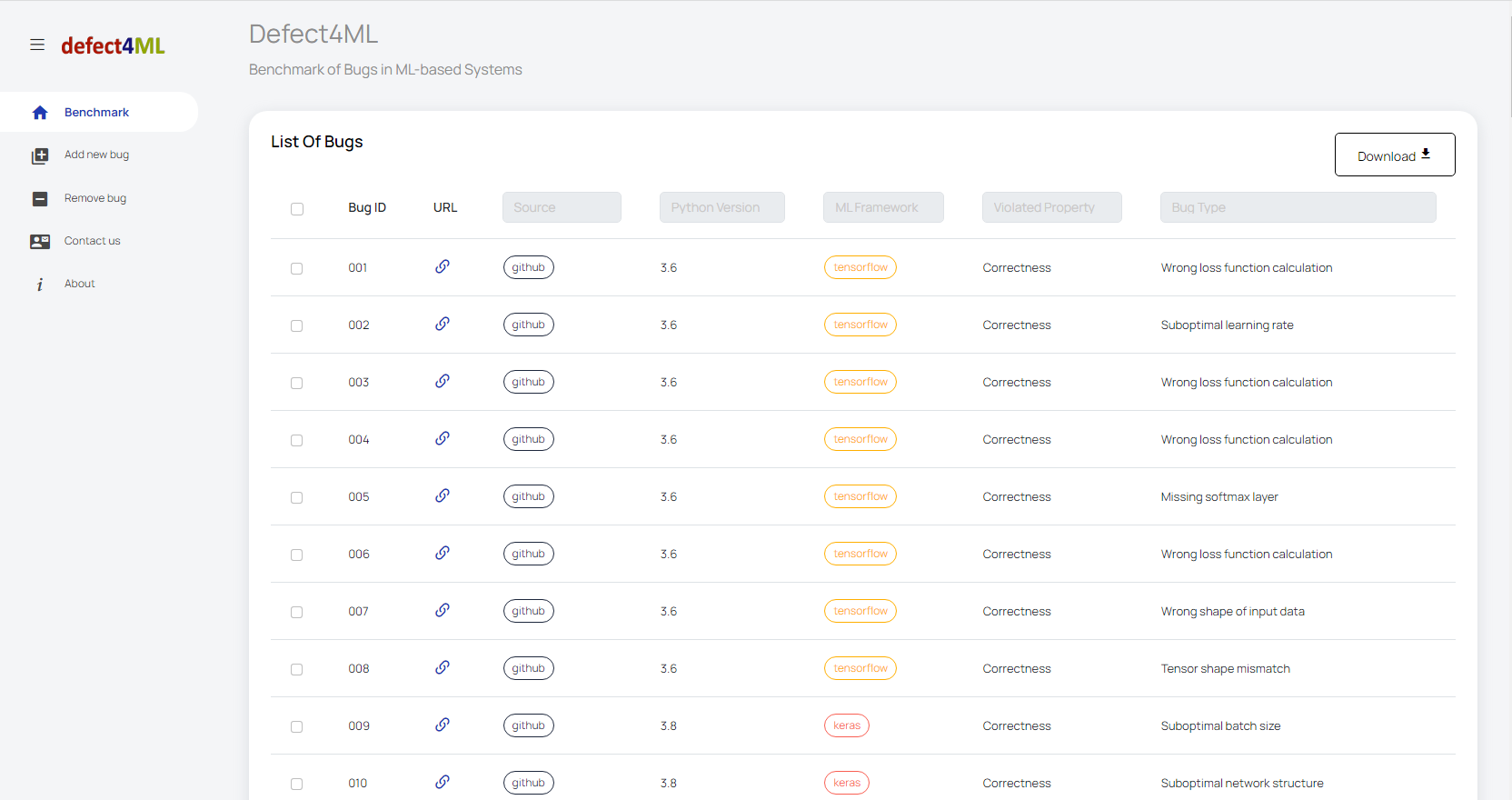}
    \caption{A screenshot of \benchmarktool{} web application.}
    \label{fig:web_app_screenshot}
    \vspace{-2em}
\end{figure}

\subsubsection{Provided API
}
In order to deliver an easy-to-use benchmark, we provide a web application endpoint for \benchmarktool~ (accessible via ~\url{http://defect4aitesting.soccerlab.polymtl.ca}). Figure~\ref{fig:web_app_screenshot} represents a screenshot of the \benchmarktool{} web application. 
Users can browse the bugs, and filter them based on different criteria 
to attain a list of bugs that is suitable for their goals. 
Since \benchmarktool~ is in the process of expansion,
we have provided the possibility for the users to add new bugs to the benchmark or raise a request for removing the existing ones. Users can submit new bugs by providing an 
artifact of the bug. Detailed explanation of adding new bugs has been presented in benchmark web application page.

\section{Discussion}
\label{sec:discussion}
This section presents an example application 
of \benchmarktool~as a case study: comparing two ML-based systems testing tools: \textit{NeuraLint} and \textit{DeepLocalize}.

\subsection{Benchmark Applications}
According to the characteristics of the \benchmarktool, it can be beneficial to studies on bugs in ML-based systems. Developers of ML-based systems testing tools can use \benchmarktool{} to show the advantages of their proposed tools and techniques compared to the existing ones. The researchers can also use our proposed benchmark to evaluate existing ML-based systems testing tools and clarify the most critical challenges that should receive attention from new studies.
We provide a case study of using \benchmarktool{} to evaluate ML-based systems testing tools. So, the primary goals of this case study are the assessment of ML-based systems testing tools and comparing them. 

In this case study, we compare two testing tools for ML-based systems. We selected two up-to-date testing tools published recently: \textit{NeuraLint}~\cite{nikanjam2021automatic} and \textit{DeepLocalize}~\cite{wardat2021deeplocalize}. \textit{NeuraLint} is a model-based automatic fault detection tools for DL programs. The authors proposed a meta-model for DL programs that contains their basic properties. To detect the bugs, \textit{NeuraLint} first extracts the graph of the DL program from its code. In the next step, it identifies the bugs using graph transformations that represent detection rules.
It is worth noting that \textit{NeuraLint} is based on static analysis of DL programs meaning that it does not need to run the DL program to identify the bugs.  

\textit{DeepLocalize} is another testing tool that can analyze DL programs, detect the bugs, and localize them automatically. It provides a customized callback function for \textit{Keras} that collects DNN detailed information during the training process. In other words,  \textit{DeepLocalize} analyses the DNN training traces to identify the possible bugs and their root causes (e.g., faulty layers or hyperparameters). Unlike \textit{NeuraLint} that analyzes the DL programs statically, \textit{DeepLocalize} uses dynamic analysis. That is, DL programs should be executable without any compilation error to be analyzable by \textit{DeepLocalize}. 

Concerning the fact that \textit{NeuraLint} can analyze the DL programs that have been written in one file, 
we had a limited number of bugs to use. On the other hand, due to the existence of compile-time errors in the buggy version of some bugs, they are not usable for \textit{DeepLocalize}.

We selected 20 bugs randomly from \benchmarktool, 10 from SO bugs and 10 from GitHub based ones. With respect to the limitations of the \textit{NeuraLint} and \textit{DeepLocalize} tools, we could just use 12 out of 20 which are usable for at least one of the tools. Table 6 demonstrates the result of evaluating \textit{NeuraLint} and \textit{DeepLocalize}. The cells filled with \textbf{$\times$} refer to the samples that have compile-time errors and could not be used by \textit{DeepLocalize}.

Table~\ref{tab:casestudies_result} demonstrates the result of evaluating \textit{NeuraLint} and \textit{DeepLocalize}. The cells filled with $\times$ refer to the samples that have compile-time errors and could not be used by \textit{DeepLocalize}. Based on the gathered results, \textit{NeuraLint} was able to identify bugs in 9 out of 10 samples. Besides, it has detected design issues in some examples, in addition to the reported bugs. Design issues are poor design and/or configuration decisions that can have a negative impact on the performance and then quality of a DL-based software system~\cite{nikanjam2021smell,nikanjam2021automatic}. For instance, bug \#91 is related to the wrong activation function of the DL model, while \textit{NeuraLint} has also identified that window size for spatial filtering does not define properly. Conversely, \textit{DeepLocalize} could localize the bug correctly in 1 out of 4 samples.

\begin{table*}[]
 \caption{
 Results of studied tools as case study}
\scriptsize	
    \centering
    \scriptsize	
    \begin{tabular}{||p{0.5cm}|p{0.8cm}|p{1.4cm}|p{1.3cm}|p{1.2cm}|p{1.6cm}|p{1.8cm}||}
        \hline
            \textbf{Bug ID} & \textbf{Source} & \textbf{Framework} & \textbf{Violated Property} & \textbf{Bug Type} & \textbf{NeuraLint result}&\textbf{DeepLocalize result}\\
            \hline\hline
            25 & GitHub & Keras & Correctness  &Wrong network architecture &  no identified error & batch 4 layer 9 : error in forward \\
            \hline
            26 & GitHub & Keras &Correctness  & Wrong type of activation function & no identified error & batch 4 layer 11 : error in forward \\
            \hline
            44 & GitHub & Keras & Efficiency &Suboptimal network structure  & no identified error & batch 4 layer 12 : error in forward \\
            \hline
            80 & SO & Keras &Correctness & Mising flatten layer & lack of pooling, missing flatten & \multicolumn{1}{c||}{\textbf{$\times$}}\\
            \hline
            84 & SO & Keras & Correctness &  Wrong type of activation function& wrong activation function, wrong units' shape & batch 0 layer 2 : error in delta weights \\
            \hline
             86 & SO & Keras &Correctness  & Wrong type of activation function  & wrong activation function, wrong layers' structure & batch 0 layer 0 : error in forward \\
             \hline
             88 & SO & Keras & Correctness &Wrong type of activation function & wrong activation function & \multicolumn{1}{c||}{\textbf{$\times$}} \\
             \hline
             89 & SO & Keras & Correctness &Wrong type of activation function & wrong activation function  & batch 0 layer 0 : error in forward\\
             \hline
             92 & SO & Keras & Correctness &  Wrong type of activation function& wrong activation function, wrong window size for spatial filtering & batch 0 layer 9 : error in delta weights \\
             \hline
              95 & SO & Keras & Correctness & Wrong API usage & no identified error & model does not learn \\
             \hline
             111 & SO & Keras & Correctness &  Missing preprocessing & no identified error & Error in delta weights \\
             \hline
             112 & SO & Keras & Correctness &  Missing flatten layer & missing flatten &  \multicolumn{1}{c||}{\textbf{$\times$}} \\
             \hline
    \end{tabular}
       \label{tab:casestudies_result}
\end{table*}

\section{Related Work}
\label{sec:related-work}
The closest work to our proposed benchmark was carried out by Kim et al.~\cite{kim2021denchmark} providing a benchmark of bugs in ML-based systems, which is called \textit{Denchmark}. They extracted 4577 bugs reported in the issue tracker of 193 GitHub repositories. Although their benchmark was the first bug benchmark focused on ML-based systems, their study has several shortcomings. Firstly, they have considered repositories with various programming languages such as Java, C, C++, Python, etc., without considering any categorization on them. So, developers might have to inspect and then categorize the bugs based on their favorite programming languages, which can be time-consuming. 
The second drawback is ignorance of the big difference between bugs  related to the ML (ML-related bugs) and other ones. \textit{Denchmark} has reported all bugs without any differentiation. 
Moreover, this study has taken no notice of bug reproducibility, which is one of the main challenges in the SRE of ML-based systems~\cite{zhang2018empirical}. Last but not least, this benchmark has neglected standard benchmark criteria, which may result in benchmark effectiveness detriment. 

Wardat et al.~\cite{wardat2021deeplocalize} also provided a benchmark of 40 bugs to validate their proposed tool. They offered 11 bugs from GitHub and 29 bugs from SO and introduced them as a benchmark of bugs in ML-based systems. The first concern about their proposed benchmark is ignorance of the standard benchmark criteria (e.g., relevance, reproducibility, fairness) and SRE challenges in ML-based systems. Besides, no information has been provided about the execution process of applications extracted from GitHub to trigger the bug.

\section{Threats to Validity}
\label{sec:validity}
We now discuss the threats to validity of our study.
\subsection{Internal Validity}

The primary source of internal threat to the validity of the results provided in this study can be the categorization of bugs. To diminish this threat, we used the predefined taxonomy of bugs in DL programs discussed in~\cite{humbatova2020taxonomy}. Another internal threat to our proposed benchmark can be the manual inspection of the bugs and making decisions about their inclusion or exclusion. The first author (a PhD student and practitioner of ML development) reviewed 100 bug-fix commits and discussed the result with the second and third authors (two PhDs with research background in engineering ML-based systems) to mitigate this threat. After three meetings, we reached an agreement on the including and excluding rules of filtering the bug-fix commits. To ensure that the benchmark consists of real bugs, we used just the SO posts with an accepted answer and bug-fix commits that clearly explain the bug and its symptoms. Verifiability of the bug fixes may be the last internal threat to the validity of this research. To counteract this threat, we used different methods for ML-related bugs extracted from GitHub and SO.

Regarding the bugs extracted from GitHub, if the buggy applications do not provide appropriate tests to verify the bug fix, we used bug fix commit messages that clearly mention the changes to fix the reported bug. To verify bugs gathered from SO posts, we only used the posts with an accepted answer that is considered as evidence of fixing the bug correctly.

\subsection{External Validity}

The most crucial threat to the external validity of this study is its limitation to the \textit{TensorFlow}~\cite{abadi2016tensorflow} and \textit{Keras}~\cite{chollet2018keras} frameworks. Firstly, we have selected them because they are two of the most popular ML frameworks~\cite{top_ml_frameworks,humbatova2020taxonomy}. Second, we have provided the feature to add new bugs using any ML framework to the benchmark and request to remove the existing bugs from it by any user. Furthermore, to achieve the highest diversity of ML-related bugs, we did not use any popularity-based filter on the GitHub repositories. We collected the bugs from the repositories developed by users with various expertise levels. Another thread to the external validity of this study can be the selection of \textit{Python} as programming language of ML-based systems development. The main reason behind the selection of \textit{Python} is the fact that it is the most used programming language for developing ML components~\cite{voskoglou2017best,Gupta:MLLangugae}. As a result, a larger variety of ML-related bugs can be found in ML-based systems based on the \textit{Python} programming language and defect4ML will be also usable for a higher number of users.

\section{Conclusion and Future Works}
\label{sec:conclusion}
The growing tendency to apply ML-based systems in safety-critical areas increases the demand for reliable ML-based systems. 
A benchmark of bugs in ML-based systems, a faultload benchmark, is a key requirement for assessing the effectiveness of studies on ML-based systems' reliability (like testing) which are based on bugs' lifecycle. 
In this study, we reviewed 1777 ML-related bugs from GitHub and 1296 from SO which are related to the ML-based systems using two of the most common ML frameworks, \textit{TensorFlow} and \textit{Keras}, and represented that only near 3.48\% of GitHub bugs and 2.93\% of reported bugs in SO are reproducible.
Besides, we showed that almost 13.3\% of fixing of all reproducible bugs extracted from GitHub can be verified by their provided test cases. However, none of the SO posts has test cases for verifying bug fixes. 
We have also proposed \benchmarktool, a faultload benchmark of ML-based systems consisting of 100 bugs extracted from the software systems using \textit{TensorFlow} and/or \textit{Keras} (62 from Github, and 38 from SO).
All of the standardized benchmark criteria have been satisfied by our proposed benchmark. 
\benchmarktool~ also addresses the main SRE challenges in ML-based systems by providing bugs in various ML frameworks with different versions, comprehensive information regarding dependencies and required data to trigger the bug, detailed information about the type of bugs, and link to the origin of the bug. 
Concerning the ongoing nature of creating benchmarks, we plan to add more bugs to cover all types of bugs in ML-based systems. 
Moreover, we are going to improve \benchmarktool{} to be usable for automatic bug repair tools. 
Besides, we will add bugs based on other ML frameworks (such as \textit{PyTorch}) to improve the coverage of the \benchmarktool. 

\bibliographystyle{spmpsci}
\bibliography{bibliography}

\end{document}